\documentclass[aps,preprint,showpacs,amsmath,amssymb]{revtex4}
\usepackage{graphicx,dcolumn,bm,bbm,amssymb}
\begin{document}

{\small [R. D. Vilela and A. E. Motter, Phys. Rev. Lett. {\bf 99}, 264101 (2007)]}\\

\centerline{\bf\large Can aerosols be trapped in open flows?}

\vspace{0.4cm}

\centerline{Rafael D. Vilela$^1$ and Adilson E. Motter$^2$}

\baselineskip 14pt

\centerline{\it\small $^1$Max Planck Institute for the Physics of Complex Systems, 01187 Dresden, Germany}

\centerline{\it\small $^2$Department of Physics and Astronomy, Northwestern University, Evanston, IL 60208, USA}

%\pacs{05.45.-a, 47.52.+j, 47.55.Kf}

\vspace{0.8cm}

\noindent
{\bf 
The fate of aerosols in open flows is relevant
in a variety of physical contexts.
Previous results are consistent with the assumption
that such finite-size particles
always escape in open chaotic advection. 
Here we show that a different behavior is possible.
We analyze the dynamics of aerosols
both in the absence and presence
of gravitational effects, and both when
the dynamics of the fluid particles is hyperbolic and nonhyperbolic.
Permanent trapping of aerosols
much heavier than the advecting fluid
is shown to occur in all these 
cases.
This phenomenon is determined by the occurrence of multiple 
vortices in the flow and is 
predicted
to happen 
for realistic particle-fluid density ratios.}

\vspace{0.5cm}

In open chaotic advection \cite{openchaoticadvection}, 
fluid particles are injected into some domain 
containing a chaotic saddle \cite{chaoSaddle}, 
where they move chaotically for a finite amount of time  
until they finally escape.
This is the history of almost all 
the injected particles in the typical case
where the fluid is incompressible. 
When instead {\it finite-size} particles are injected into the domain, 
the outcome can be fundamentally different \cite{ott}. 
The reason is that these particles are subjected to 
new forces, in particular the Stokes drag.
Their dynamics is  dissipative, 
allowing the existence of attractors where the particles
can be trapped.
This has been shown to be typical in the case of {\it bubbles}, 
i.e., finite-size particles less dense 
than the fluid \cite{benczik}.
On the other hand, 
previous studies on specific
open flows support the assumption that 
finite-size particles denser than the
fluid, called {\it aerosols}, always escape \cite{benczik,motterPREnov2003}.
Moreover, there is evidence that aerosols escape even faster
than the particles of fluid \cite{benczik}.
The physical reason for this 
is that 
chaotic saddles are usually associated with 
the occurrence of vortices. 
Bubbles in vortices tend to move inwards,
therefore possibly being trapped,
whereas aerosols tend to 
spiral
outwards, therefore escaping and possibly doing so faster
than the fluid itself.

The dynamics of heavy particles in open flows is important in many fields.
In astrophysics, it can provide a mechanism for the formation of planetesimals
in the primordial 
solar nebula \cite{barge;tanga}. In
geophysics,
it has direct consequences for the transport and activity of pollutants and cloud
droplets in the atmosphere \cite{atmos}. It can also be useful for particle 
separation in industrial applications.

In this Letter, we show that  
the premise that aerosols in vortices tend to
move outwards does not necessarily imply
that these particles will always escape faster in open flows.
More remarkably, we show that aerosols 
much heavier than the advecting fluid
can be 
{\it permanently trapped} in open chaotic advection.
The mechanism affording the trapping of
aerosols is 
associated with the occurrence of two or more 
vortices in a bounded region of the flow.
In escaping from a vortex, the aerosols may 
enter the domain of another vortex, which in its
turn may drive the particles back to the first vortex.
This 
can
lead to the formation of bounded stable orbits
that give rise to an attractor. 
We illustrate this phenomenon for two widely studied open flows:
the {\it blinking vortex
system}, which has static vortices,
and the {\it leapfrogging vortex system}, which has moving vortices.
These systems do not include any physical obstacle 
or boundary layer effect \cite{marchioli}
that
could mask the purely dynamical phenomenon 
we are interested in.
The trapping of aerosols in open flows
sharply contrasts with the behavior of the
fluid particles, which escape and go to infinity
with probability 1.

% ----------------------------------------------

We start with the 
equation of motion for a small heavy spherical particle
in a fluid flow. In dimensionless form, it
reads \cite{maxey}
\begin{equation}
\ddot{\bf r}=A\left({\bf u}-\dot{\bf r}-W{\bf n}\right),
\label{maxey}
\end{equation}
where 
${\bf r}$ is the position vector of the particle, 
${\bf u}={\bf u}({\bf r}(t),t)$ is the fluid velocity field
evaluated at the particle's position, and
${\bf n}$ is a unit vector pointing upwards in the vertical direction.
The two parameters governing the dynamics are the
inertia parameter $A$ and the gravitational parameter $W$.
They can be written in terms  of the 
characteristic
length $L$ and velocity $U$ of the flow,
radius $a$ of the particle, 
kinematic viscosity $\nu$ of the fluid,
gravitational acceleration $g$, and 
densities  $\rho_p$
and $\rho_f$
of particle and fluid, respectively.
The defining equations for these parameters are
${W}=(2a^2 \rho_p g)/(9\nu U\rho_f)$ and
$A={R}/{St}$, where
$R = {\rho_f}/{\rho_p} \ll 1$
and $St=(2a^2U)/(9\nu L)$ is the Stokes number of the particle.
In the case of a water droplet in air flow, for example, 
one has
$\rho_f / \rho_p \sim 10^{-3}$.

We first consider the blinking vortex-source system \cite{arefkarolyi}.
This  2-dimensional system is periodic in time and consists of two
alternately 
point sources in a plane.
It models the alternate injection of rotating fluid
in a large shallow basin.
The blinking vortex-source system is described by the streamfunction
\begin{equation}
\Psi=-\left(K\ln r'+Q\phi'\right) \Theta(\tau)
-\left(K\ln r''+Q\phi''\right) \Theta(-\tau),
\label{stream}
\end{equation}
where $\Theta$ stands for the Heaviside step function, 
$\tau=0.5 T - (t \mbox{ mod } T)$, 
and $T$ is the period of the flow.
Here, $r'$ and $\phi'$ are
polar coordinates centered at $(x,y)=(-1,0)$, 
while 
$r''$ and $\phi''$ are
polar coordinates centered at $(1,0)$.
The parameters $Q$ and $K$ are the strengths of the
source and vortex, respectively. 
The parameter $Q$ is 
negative (in fact, 
positive values of $Q$ define a  
vortex-sink system).
Positive and negative values of $K$ correspond, respectively, to 
counterclockwise and clockwise vortex motion. 
The sources are located at positions $(\pm 1,0)$.
For each half-period, the flow remains steady with only
one of the sources open. 
In the time interval $0 < t \mbox{ mod } T < 0.5T$, the open source
is the one at $(-1,0)$, whereas  
in the time interval $0.5T < t \mbox{ mod } T < T$, the open source
is the one at $(1,0)$.
The velocity field of the  
fluid flow is
\begin{equation}
{\bf u}=(u_x , u_y ), \hspace{3mm} u_x=\partial \Psi /\partial y,
\hspace{3mm} u_y=-\partial \Psi /\partial x.
\label{fluid}
\end{equation}

The parameters $Q$, $K$, and $T$ in Eq. (\ref{stream})
can be chosen  
to yield either a nonhyperbolic
or a hyperbolic dynamics for the fluid particles 
({\it passive advection}).
For instance, for $Q=-20$, $K=-400$, and $T=0.1$, 
the invariant set is nonhyperbolic as it consists of a chaotic
saddle plus (at least) one KAM island.
For $Q=-10$, $K=-160$, and $T=0.1$, on the other hand, apparently there are no islands.
Numerical calculations confirm that the survival probability of fluid particles in
the neighborhood of the chaotic saddle decays exponentially fast, which is a signature of
hyperbolic dynamics.

% ------------------------FIGURE 1---------------------------

\begin{figure}[tbh]
\includegraphics[angle=0,width=8.0cm,height=7.0cm]{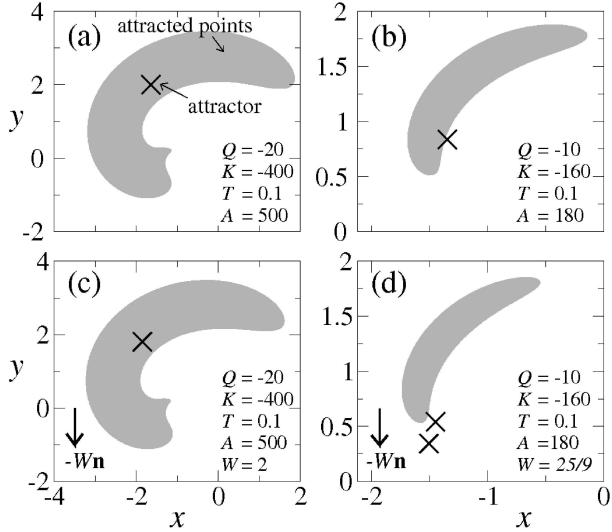}
\caption{ \baselineskip 14pt
Projection into the physical space of the 
stroboscopic 
section $t \mbox{ mod } T =0$ 
for the aerosol dynamics in the flow of Eq. (\ref{stream}).
The black $\times$ symbols indicate the attractors  
and the gray areas 
the corresponding basins 
of attraction for initial velocities equal to the local velocities
of the fluid at $t=0$.
(a) Nonhyperbolic 
and (b) hyperbolic passive advection in the absence of gravity;
(c) nonhyperbolic 
and (d) hyperbolic passive advection in the presence of gravity.
The attractors in (a-c) are period-one orbits,
whereas the attractor in (d) is a period-two orbit.
Note that the attractors are not necessarily inside the gray
areas because these areas correspond to {\it subsets} of 
the basins of attraction defined by specific initial velocities.
The gravity vector points
in the negative direction of the y-axis.
\label{fig2}}
\end{figure}

% ------------------------ END FIGURE 1---------------------------

We now investigate the dynamics of aerosols
governed by Eq. (\ref{maxey}) in the
flow given by Eqs.
(\ref{stream}) and (\ref{fluid}),
both in the cases where the passive advection is hyperbolic and 
nonhyperbolic, 
and both in the absence and presence of gravity.
As shown in Fig.~\ref{fig2}, the trapping of aerosols 
in attractors can occur
in all these cases.
The figure shows, for different parameters,  
the projection into the physical space
of the  attractor at time instants $t \mbox{ mod } T =0$. 
The corresponding basins of attraction 
for initial velocities equal to the local velocities of the 
fluid are also shown.
We note that the dynamics of the aerosols 
takes place
in a 4-dimensional phase space, corresponding 
to the variables $(x,y,v_x,v_y)$, and 
is 
dissipative due to the drag term,  whereas the 
dynamics of the fluid particles is 2-dimensional,  
since their velocity is a function of their position, 
and is conservative because the flow is incompressible. 
Despite the fact that, from the dynamical systems viewpoint,    
dissipation 
may give rise to attractors, 
we stress that we deal with {\it open} flows
and that in this case the presence of dissipation {\it does not} imply 
the occurrence of attractors  in a {\it bounded region}.
In open systems, attractors may be formed at {\it infinity} in the physical
space \cite{motterPREnov2003}.
The trapping of aerosols is counterintuitive also in view of 
the fact that these particles move outwards in vortices.
Notwithstanding, Fig.~\ref{fig2} shows that this phenomenon 
is possible in a broad range of conditions.
Figure \ref{fig2}(a) shows the occurrence of attractors for the dynamics of the aerosols 
when the underlying passive advection is nonhyperbolic. 
For small $W$ and large $A$, the dynamics of the aerosols
can be understood as a perturbation of the dynamics of the fluid particles 
\cite{maxey87}. 
As shown below, in this 
regime attractors can be formed in
the KAM islands of the passive advection.
In the case of Fig.~\ref{fig2}(a), we are not in the limit of weak perturbation 
and the basin of attraction is actually larger than the KAM island itself.  
The appearance of attractors when  the aerosols are advected by a 
{\it hyperbolic} passive 
dynamics
is less expected,
since hyperbolic systems are structurally stable.
Nevertheless, the occurrence of attractors is possible also in this case, 
as shown in Fig.~\ref{fig2}(b), and the reason again is that   
we are far from the weak perturbation limit.  
Attractors also occur  
when the gravitational effects are important, i.e., when $W$ is of 
order of 1 or larger. As shown in 
Figs.~\ref{fig2}(c) and \ref{fig2}(d), this is possible
for both nonhyperbolic and hyperbolic passive advection.
We consider that the gravitational field points along the 
   $y$-direction. 

It is worth noting the 
rich variety of possibilities for
the motion of aerosols under gravity, which
includes both the trapping in smooth {\it open} flows described here,
the suspension in random {\it closed} flows discussed in \cite{pasquero},  
and the increased average 
settling velocity when the aerosols are in an infinite, periodic,  
cellular fluid flow \cite{maxey86}.
Note that the mechanism for suspension presented in \cite{pasquero}
is based on the change of sign of the curvature of the streamlines
while our mechanism is based on the vortex-to-vortex advection of the aerosols.
In particular, the curvature of the streamlines of the blinking vortex-source system
is one-signed and does not satisfy the conditions considered in \cite{pasquero}.

\begin{figure}[t] %th
\includegraphics[width=0.45\textwidth]{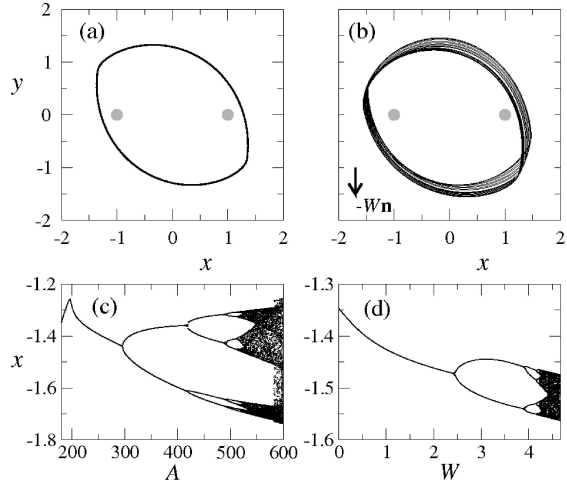} 
\caption{ \baselineskip 14pt
(a) Periodic attractor corresponding to 
$W=0$ and 
(b) strange attractor 
corresponding to 
$W=40/9$.
The gray dots represent the point sources.
Bifurcation diagrams showing the $x$-coordinates
of the attractor at time $t \mbox{ mod } T =0$
as a function of parameter $A$
in the absence of gravity (c) and  
as a function of $W$ when $A=180$ (d).
The unspecified parameters are the same as in Fig.~\ref{fig2}(b).
\label{fig3}}
\end{figure}

The stroboscopic sections shown in Fig.~\ref{fig2} 
correspond to attractors that 
are simple periodic orbits. 
The full physical space projection of one 
such attractor is shown in Fig.~\ref{fig3}(a).  
But strange attractors can also occur, as shown in Fig.~\ref{fig3}(b). 
They are formed when 
either $A$ or $W$ is increased (see Figs.~\ref{fig3}(c-d)).
The bifurcation diagrams are dominated %, in both cases, 
by the period-doubling route.

% -------------------------FIGURE 2--------------------------

% -------------------------END FIGURE 2--------------------------

% -------------------------FIGURE 3--------------------------

\begin{figure}[tbh]
\includegraphics[width=0.45\textwidth]{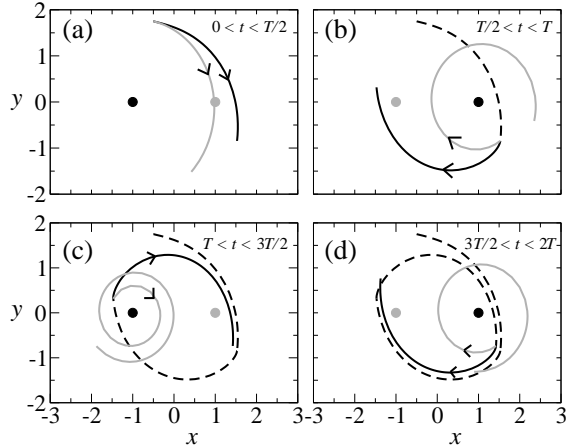}
\caption{ \baselineskip 14pt
Trajectories of an aerosol (black)
and a fluid particle (gray)
of same initial position.
In (a), the aerosol and the fluid particle also have the
same initial velocity, whereas
in (b-d) the aerosol simply continues the trajectory 
initiated in (a) and detaches from the fluid.
The point source that is open during each time interval
is shown as a black dot (the gray dot corresponds to the other one).
The dashed lines represent
the trajectory of the aerosol
before the time interval indicated in the figure. 
After 2 periods of the fluid flow, 
the aerosol is already very close to 
a periodic attractor [cf. Fig.~\ref{fig3}(a)]. 
The parameters are the same as in Fig.~\ref{fig2}(b).
\label{fig5}}
\end{figure}

% -------------------------END FIGURE 3--------------------------

The trapping of aerosols in open flows displaying chaotic advection is a general
phenomenon. The dynamics defined by Eq.~(\ref{maxey}) is dissipative. In open
flows, dissipation is not a sufficient condition for the occurrence of  
bounded attractors.
However, this condition becomes sufficient 
in periodic flows 
if a
set of initial conditions with positive volume is advected back to itself after 
one time period.
In the case of bubbles, this condition is frequently satisfied because bubbles 
tend to remain inside closed orbits of the 
fluid particles generated by vortices \cite{benczik,motterPREnov2003}.
This mechanism cannot explain the existence of attractors in the case of
aerosols because, for heavy particles, rather the opposite happens due to
the centrifugal force. However, when {\it multiple} vortices are present 
in the flow and remain confined to a bounded region, 
we show that a {\it different} mechanism can give rise to attractors 
in which the aerosols are trapped. 
The mechanism of trapping is in this case based on successive ``escape attempts" 
from distinct vortices. As shown in Fig.~\ref{fig5}, a possible outcome of the motion of
aerosols outwards  successive vortices is the formation of bounded orbits.
Trapping occurs if this happens for
a set of orbits with positive volume, as shown 
for the blinking vortex system.

\begin{figure}[t]
\includegraphics[width=0.45\textwidth]{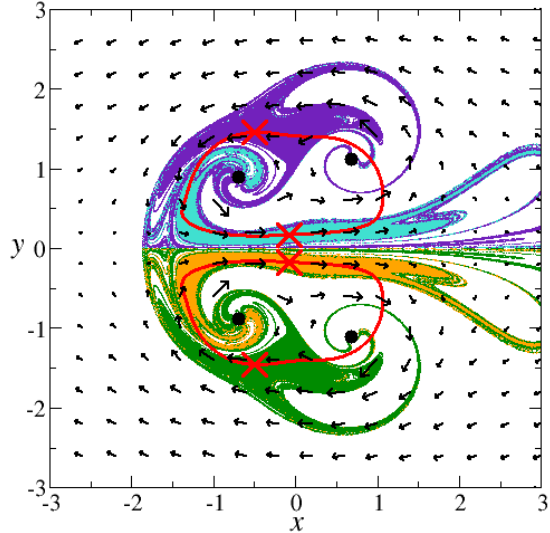}
\caption{  \baselineskip 14pt %(color online)
Trapping of aerosols in the leapfrogging vortex flow: physical space projection of 
the attractors (red $\times$ symbols) and corresponding basins of attraction 
(colored regions) for initial velocities equal to the fluid velocity at $t$ mod
$T$ $=0.8$, where $T$ is the period of the flow. Also shown are the
velocity field of the fluid (black arrows) and the positions of the vortices 
(black dots) at the same instant. The red curves indicate the orbits described
by the attractors. See animation at 
http://www.pks.mpg.de/$\sim$rdvilela/leapfrogging.html
\label{fig6}}
\end{figure}

To demonstrate the generality of our results, we now consider a system
with moving vortices: the leapfrogging vortex system \cite{leap},
which is a 2-dimensional flow consisting of two vortex pairs of equal strenghts.
These vortex pairs move along the direction of
a symmetry axis that separates them as a consequence
of mutual influence \cite{arefprovenzale}.
Denoting the positions of the vortices by $(x_1,\pm y_1)$ and $(x_2,\pm y_2)$,
we take $x_1 =x_2 =0$, $y_1 =0.5$, and $y_2 =1.5$ as the vortices coordinates
at $t=0$. The Hamiltonian 
$H(x_0,x_r,2y_0,\frac{y_r}{2})=0.5\ln \frac{(x_r^2 +(2y_0 )^2)((2y_0 )^2-4(y_r /2)^2 )}
{(x_r^2 +4(y_r /2)^2 )}$ 
describes the periodic motion of the vortices, 
where $x_0 =(x_1 +x_2 )/2$, $y_0 =(y_1 +y_2 )/2$, 
$x_r =(x_2 -x_1 )$, and $y_r =(y_2 -y_1 )$.
The streamfunction in a reference frame whose origin
is the point $(x_0(t),0)$
reads $\Psi(x,y,t)=\ln [(r_3 r_4 )/(r_1 r_2 )]-\dot{x}_0 (t)y$, where 
$r_{1(2)}^2 =[x-x_{1(2)} (t)]^2 + [y-y_{1(2)} (t)]^2$ 
and 
$r_{3(4)}^2 =[x-x_{2(1)} (t)]^2 + [y+y_{2(1)} (t)]^2$.
In this reference frame, the fluid particles come from
$x=\infty$, are scattered in the region close to the origin, where 
the vortices are confined, 
and finally move towards the negative direction of the $x$-axis.
We 
%investigate 
have investigated
the dynamics of aerosols in this flow
and 
%find 
we found
trapping for a broad range of the parameter $A$.
Figure \ref{fig6} illustrates the trapping of aerosols in this flow for $A=50$ and $W=0$.

% -------------------------FIGURE 4--------------------------

% -------------------------END FIGURE 4--------------------------

In the general case of nonhyperbolic passive advection, we can 
demonstrate the formation of attractors explicitly for $A \gg 1$ and $W\ll 1$
using a first-order approximation \cite{maxey87} of the dynamics given by $\dot{\bf r}=
{\bf u}-W{\bf n}-\frac{1}{A}[\frac{\partial {\bf u}} {\partial t}+({\bf u} \cdot \nabla)
{\bf u} - (W{\bf n} \cdot \nabla) {\bf u} ]$. If $W\ll 1$ and the magnitude of the term
inside the brackets is much smaller than $A$, the dynamics corresponds to a  small 
perturbation of the passive advection in a $2$-dimensional effective phase space.
If the divergence   $\nabla \cdot \dot{\bf r}$ of the velocity field is negative,
as in the blinking vortex-source system \cite{obs}, KAM islands of the passive advection
are expected to be transformed into basins of attraction. The formation of attractors in 
KAM islands has been observed in the advection of bubbles \cite{motterPREnov2003} 
and in the study of Hamiltonian 
systems \cite{feudel}. However, in a very neat contrast with the mechanism previously considered 
in the study of bubbles, in the case of aerosols the KAM islands that give rise to 
attractors describe the dynamics of particles 
that {\it necessarily visit more than one vortex}.
From the relation $\nabla \cdot \dot{\bf r}= (\omega^2-s^2)/(2A)$ \cite{maxey87}, we can
see that in these islands the strain $s$ must dominate over the vorticity $\omega$.
We emphasize, however, that the phenomenon of trapping of aerosols is far more general
since it also occurs in non-perturbative regimes (small $A$ and large $W$) and in the
absence of KAM islands, as shown in Figs.~\ref{fig2}-\ref{fig6}.

In conclusion, we have shown the 
occurrence of 
trapping of heavy particles in open flows.
This phenomenon does not
depend on the nonhyperbolicity of the passive advection 
and is possible even
when the gravitational effect is large.
An experiment to demonstrate trapping 
in the leapfrogging vortex system
could be done with air as the working fluid \cite{yamada}.
In this case, the condition
${\rho_f}/{\rho_p} \ll 1$ is fulfilled for virtually all solid 
and liquid particles.
The trapping mechanism reported 
%herein provides the concentration
%of heavy particles in %{\bf point-like} 
%attractors. 
here provides a mechanism for the concentration of heavy particles in
specific regions of the physical space.
This may play a role in planetesimal formation
in primordial nebula with rapidly rotating anticyclonic vortices, 
where the centrifugal force prevails over the Coriolis force.
It may 
%be also 
also be
useful for particle-fluid and particle-particle 
(different size classes)
separation 
%using continuous flow 
in industrial applications.\\

%\begin{acknowledgments}
We thank Ernesto Nicola, Edward Ott, 
Oreste Piro, Antonello Provenzale, and Tam\'as T\'el for 
stimulating
discussions.
%\end{acknowledgments} 

\end{document}